\documentclass[11pt,a4paper]{article}

\usepackage[utf8]{inputenc}
\usepackage[T1]{fontenc}
\usepackage[english]{babel}

\usepackage[
a4paper,
top=1in,
bottom=1in,
left=1in,
right=1in
]{geometry}

\usepackage[protrusion=true,expansion=true]{microtype}

\usepackage{setspace}
\usepackage{amsmath}
\usepackage{amssymb}
\usepackage{amsthm}
\usepackage{bm}

\newcommand{\E}{\mathbb{E}}

\newcommand{\calS}{\mathcal{S}}

\usepackage{booktabs}
\usepackage{longtable}
\usepackage{array}
\usepackage{multirow}

\usepackage{graphicx}
\usepackage{caption}
\usepackage[
round,
authoryear,
sort&compress
]{natbib}

\usepackage{titlesec}

\titleformat{\section}
{\large\bfseries}
{\thesection.}
{0.5em}
{}

\titleformat{\subsection}
{\normalsize\bfseries}
{\thesubsection}
{0.5em}
{}

\titleformat{\subsubsection}
{\normalsize\bfseries}
{\thesubsubsection}
{0.5em}
{}

\usepackage{hyperref}

\hypersetup{
colorlinks=false,
pdfborder={0 0 0}
}

\usepackage{xcolor}
\usepackage{float}

\title{\textbf{Multi-Scale Markov-Switching GARCH}:\\ Volatility Regime Detection in EUR/USD}
\author{Jayesh Chaudhary\\Independent Researcher}
\date{\textbf{May, 2026}}

\begin{document}
\maketitle
\begin{abstract}
Financial time series exhibit pronounced non-stationarity, rendering single-regime volatility models structurally misspecified for foreign exchange markets. This paper introduces a triple-timeframe Markov-Switching GARCH (MS-GARCH) framework with time-varying transition probabilities (TVTP) for volatility regime detection in EUR/USD, operating simultaneously across daily (1D Macro), four-hourly (4H Meso), and hourly (1H Micro) timescales.
Each layer uses an AR(1)-MS-GARCH model with skewed Student-t emissions and three hidden states: Calm, Turbulent, and Crisis. These states are estimated through penalized maximum likelihood using Numba-JIT compiled Hamilton filter kernels. We build a 27-dimensional joint probability tensor as the outer product of the three marginal state probability vectors, and use this tensor to route signals through a Mixture-of-Experts architecture of 27 independent RidgeCV models.

The framework is evaluated on EUR/USD data spanning 2015--2025, with a strictly out-of-sample (OOS) period of 2021--2025 comprising 31,152 hourly observations across 16 walk-forward analysis (WFA) quarters. Key empirical results are as follows. The Diebold--Mariano test confirms statistically superior volatility forecasting relative to a GARCH(1,1) benchmark: $\mathrm{DM} = +4.7040$ ($p = 1.28 \times 10^{-6}$). Kolmogorov--Smirnov distributional purity tests yield $p = 1.35 \times 10^{-153}$ (Calm vs.\ Turbulent), confirming that the three regimes represent distinct data-generating processes. The smoothed Spearman information coefficient improves from 0.5264 (GARCH) to 0.5371 (MS-GARCH). TVTP is strongly justified at the 4H and 1H scales ($\Delta\mathrm{AIC} = +690.7$ and +499.9 respectively), while the 1D model correctly employs a static specification. All three timescales preserved monotonic volatility ordering out-of-sample (Calm < Turbulent < Crisis), confirming statistically distinct regime separation. The out-of-sample directional information coefficient from the RidgeCV evaluation layer is +0.0252 ($p = 6.75 \times 10^{-5}$). The framework is positioned as a risk management tool rather than a standalone alpha engine.

\noindent \textbf{Keywords:} Markov-switching GARCH; regime detection; time-varying transition probabilities; EUR/USD; volatility forecasting; hidden Markov model; multi-scale analysis; walk-forward analysis
\end{abstract}
\tableofcontents
\newpage
%% ============================================================
\section{Introduction}
\label{sec:intro}
%% ============================================================
Financial time series are characterised by pronounced non-stationarity: the statistical properties of asset returns---their mean, variance, autocorrelation structure, and tail behaviour---shift systematically across different market environments. Central bank policy cycles, geopolitical shocks, and liquidity crises create structural breaks that render single-regime models fundamentally misspecified. The canonical GARCH(1,1) model of \citet{bollerslev1986}, while enormously influential, assumes a single data-generating process operating throughout the sample. When applied to EUR/USD returns over the 2021--2025 period, this assumption produces a notable result: the estimated ARCH coefficient $\hat{\alpha} = 0.3038$ is three to six times above the typical range of 0.05--0.10 for foreign exchange data, with $\hat{\beta} = 0.6076$ and total persistence $\hat{\alpha} + \hat{\beta} = 0.9114$. This anomalously high $\hat{\alpha}$ is consistent with model misspecification arising from the imposition of a single-regime structure on a non-stationary process. A single-regime GARCH model, confronted with a sample that contains multiple structural regimes, absorbs the regime-switching dynamics into its shock-response parameter, becoming hyper-reactive to individual shocks in order to compensate for the absence of state-dependent variance paths.

An influential solution to this problem was provided by \citet{hamilton1989}, whose Markov-switching framework introduced the concept of latent regime states governed by a first-order Markov chain. Hamilton's model demonstrated that economic time series could be parsimoniously characterised as switching between a small number of distinct data-generating processes, with the switching mechanism itself being probabilistically estimated from the data. The subsequent development of Markov-Switching GARCH (MS-GARCH) models, pioneered by \citet{gray1996} and refined by \citet{haas2004}, extended this framework to accommodate state-dependent volatility dynamics. The \citet{haas2004} specification, which runs $K$ independent parallel GARCH processes---one per regime---is particularly tractable because it avoids the path-dependence explosion that afflicts the \citet{gray1996} integrated formulation.

Despite this progress, existing MS-GARCH implementations share a common limitation: they operate at a single timescale. This is a structural misspecification for foreign exchange markets, where regime dynamics are hierarchically nested across multiple temporal scales. A central bank policy shift (a macro regime change) unfolds over months; an institutional positioning cycle (a meso regime change) unfolds over days; a liquidity-driven intraday stress event (a micro regime change) unfolds over hours. These three classes of dynamics are simultaneously present in any EUR/USD hourly return series, yet they are governed by fundamentally different data-generating processes. A single-scale HMM fitted to hourly data faces a fundamental modelling challenge: its transition matrix must simultaneously encode the very low off-diagonal probabilities appropriate for macro regime transitions ($P[\text{Calm} \to \text{Crisis}] \approx 0.001$ per day) and the higher off-diagonal probabilities appropriate for intraday microstructure transitions ($P[\text{Calm} \to \text{Turbulent}] \approx 0.08$ per hour). These requirements place conflicting demands on a single $K \times K$ transition matrix.

This paper makes three core contributions to the literature on volatility regime detection:\\

\noindent \textbf{1. Muli-Scale MS-GARCH architecture:}
We introduce a framework that fits three independent AR(1)-MS-GARCH models simultaneously at the 1D Macro, 4H Meso, and 1H Micro timescales, each optimised for the dynamics at its native scale. The outputs are combined through a 27-dimensional joint probability tensor $\mathcal{P}_t(i,j,k) = \pi_t^{(1D)}(i) \times \pi_t^{(4H)}(j) \times \pi_t^{(1H)}(k)$, enabling cross-scale interaction analysis that is not available in single-timeframe models.

\noindent \textbf{2. Time-varying transition probabilities with composite stress drivers:} We implement the \citet{filardo1994} multinomial logit TVTP specification, driven by composite microstructure stress indices constructed from volatility z-scores, spread proxies, and momentum signals. TVTP is strongly justified at the 4H and 1H scales ($\Delta\mathrm{AIC} = +690.7$ and +499.9) \citep{akaike1974}, while the 1D model correctly employs a static specification (positive AIC $= +2170.24$ suggests near-overparameterisation on 1,871 daily bars).

\noindent \textbf{3. Cross-scale probability tensor for state-dependent prediction and Shannon entropy filtering:} The joint probability tensor serves as soft routing weights for 27 independent RidgeCV expert models, each specialised for a specific (Macro, Meso, Micro) regime combination. A Shannon entropy filter \citep{shannon1948} suppresses trading during high-uncertainty environments.\\

The empirical evaluation employs EUR/USD 1-minute mid-price data from 2015 to 2025, resampled to 1H, 4H, and 1D frequencies. The out-of-sample period spans 2021-01-01 to 2025-01-01, comprising 31,152 hourly observations across 16 walk-forward analysis quarters (2021-Q1 through 2024-Q4). The WFA protocol employs a rolling 5-year training window of approximately 31,000 hourly bars, with a rolling training scheme that avoids cold-start discontinuities
across quarter boundaries.

%% ============================================================
\section{Literature Review}
\label{sec:lit}
%% ============================================================
The intellectual lineage of this paper begins with \citet{hamilton1989}, whose seminal contribution introduced the Markov-switching model as a framework for characterising non-stationary economic time series. Hamilton demonstrated that US GNP growth could be parsimoniously described as switching between two distinct data-generating processes---expansion and contraction---governed by a first-order Markov chain. The forward-backward algorithm he developed for computing filtered and smoothed state probabilities has $O(TK^{2})$ computational complexity and remains the standard inference procedure for hidden Markov models \citep{hamilton1989}. The key insight of \citet{hamilton1989} is that the latent state sequence need not be observed directly; it can be inferred probabilistically from the observable data through the recursive Hamilton filter.

The parallel development of GARCH models by \citet{bollerslev1986}, building on the ARCH framework of \citet{engle1982}, provided the standard toolkit for modelling conditional heteroskedasticity in financial time series. The GARCH(1,1) model captures two fundamental stylised facts of financial volatility: volatility clustering (periods of high volatility tend to be followed by high volatility) and mean reversion (volatility eventually returns to its unconditional level). However, as \citet{bollerslev1986} himself acknowledged, the GARCH framework assumes a single, stationary data-generating process. When applied to samples containing structural breaks, the GARCH model absorbs the non-stationarity into its parameters, producing the anomalously high ARCH coefficients documented in this paper.

The integration of Markov-switching dynamics with GARCH volatility models was first attempted by \citet{gray1996}, who proposed an MS-GARCH specification in which the conditional variance depends on the full history of regime-weighted variances. \citet{gray1996} proposed a ``collapsing'' approximation that reduces the path count at each step, but this approximation introduces bias and increases computational cost by a factor of 3--5 relative to simpler specifications.

A computationally tractable alternative was proposed by \citet{haas2004}, who proposed running $K$ independent parallel GARCH processes (one per regime) and switching between them according to the Markov chain. The \citet{haas2004} specification makes the likelihood exactly tractable because each regime's variance path depends only on the within-regime history, not on the full cross-regime history. This is the specification adopted in the present paper. The independence assumption introduces a known limitation: the within-state GARCH does not condition on volatility inherited from previous states, but this is a worthwhile trade-off for computational tractability, particularly in the multi-timeframe setting where three separate models must be estimated.

The extension of MS-GARCH to time-varying transition probabilities was pioneered by \citet{filardo1994}, who demonstrated that business cycle transition probabilities could be modelled as functions of leading economic indicators through a multinomial logit specification. \citet{filardo1994} showed that allowing transition probabilities to vary with observable covariates substantially improved the model's ability to predict regime transitions in advance, rather than merely detecting them contemporaneously. This insight is directly applicable to foreign exchange markets, where microstructure stress indices, constructed from volatility z-scores, spread proxies, and momentum signals provide genuine leading information about regime transitions.

Recent developments in the MS-GARCH literature have focused on improving both distributional flexibility and estimation methodology. Alternative heavy-tailed emission distributions, including Normal Inverse Gaussian (NIG) and Variance Gamma (VG) specifications, have been proposed to better capture extreme market movements and asymmetric tail behaviour. In parallel, Bayesian estimation methods, Markov Chain Monte Carlo techniques, and Sequential Monte Carlo approaches have been developed to address inference challenges and enable more adaptive state estimation.

Despite these advances, much of the literature continues to focus on modelling regime dynamics at a single temporal resolution. The present study extends this literature by jointly modelling regime dynamics across daily, four-hour, and hourly horizons and integrating the resulting state probabilities into a unified cross-scale probability tensor. This multi-scale architecture enables the analysis of regime interactions across temporal horizons rather than treating each timescale in isolation.

The evaluation of competing volatility forecasts requires a formal statistical test. \citet{diebold1995} developed the Diebold--Mariano (DM) test, which tests the null hypothesis of equal predictive accuracy between two forecasting models using the loss differential $d_t = L(e_{1t}) - L(e_{2t})$, where $L(\cdot)$ is a loss function (typically squared error) and $e_{it}$ is the forecast error of model $i$. Under the null, the DM statistic $\bar{d} / \sqrt{\hat{\sigma}^{2}_d / T}$ is asymptotically standard normal. A positive DM statistic indicates that model 1 has smaller average loss than model 2. The DM test is the appropriate tool for comparing the MS-GARCH and GARCH(1,1) volatility forecasts in this paper.

The probabilistic calibration of regime probability forecasts is assessed using the Brier Score of \citet{brier1950}, which measures the mean squared error of probabilistic forecasts for binary events. The Brier Skill Score (BSS) compares the model's Brier Score to that of a climatological reference forecast, with BSS $> 0$ indicating skill above the base rate. The BSS is particularly appropriate for evaluating crisis probability forecasts, where the base rate is low (2--8\% depending on the timeframe) and the cost of false negatives is high.

The residual diagnostics employ the ARCH-LM test of \citet{engle1982}, which tests for remaining conditional heteroskedasticity in the standardised residuals after model fitting. Rejection of the null hypothesis of no ARCH effects indicates that the within-state GARCH specification has not fully captured the volatility clustering dynamics, motivating extensions to higher-order GARCH or asymmetric specifications.

The Markov assumption underlying all HMM specifications implies that regime transitions depend only on the current state, not on how long the system has been in that state. \citet{durland1994} documented empirical evidence of duration dependence in US GNP growth regimes, motivating semi-Markov models in which transition probabilities depend on regime age. This limitation is acknowledged in Section~\ref{sec:limits}.

%% ============================================================
\section{Mathematical Foundations}
\label{sec:math}
%% ============================================================
\subsection{Hidden Markov Model Specification}
\label{subsec:hmm}
Let $\{S_t\}_{t=1}^T$ denote a latent Markov chain taking values in the state space $\calS = \{0, 1, 2\}$, corresponding to the Calm, Turbulent, and Crisis regimes respectively. The chain is characterised by an initial distribution $\pi_{0} = (\pi_{0}(0), \pi_{0}(1), \pi_{0}(2))^\top$ with $\sum_k \pi_{0}(k) = 1$, and a $K \times K$ transition matrix $P = [p_{ij}]$ where \begin{equation} p_{ij} = P(S_t = j \mid S_{t-1} = i), \quad \sum_{j=0}^{K-1} p_{ij} = 1 \quad \forall i.
\label{eq:markov}
\end{equation}
The first-order Markov property states that $P(S_t \mid S_{t-1}, S_{t-2}, \ldots, S_{1}) = P(S_t \mid S_{t-1}), $ so the full history of the chain is summarised by the current state. The observable return series $\{r_t\}$ is conditionally independent given the state sequence: $r_t \perp r_s \mid S_t$ for $t \neq s$.

The Hamilton filter \citep{hamilton1989} computes the filtered state probabilities $\xi_{t|t}(k) = P(S_t = k \mid r_{1}, \ldots, r_t)$ recursively. The prediction step propagates the previous filtered probabilities through the transition matrix:
\begin{equation}
\xi_{t|t-1}(j) = \sum_{k=0}^{K-1} p_{kj} \cdot \xi_{t-1|t-1}(k),
\label{eq:predict}
\end{equation}
and the update step incorporates the new observation:
\begin{equation}
\xi_{t|t}(k) = \frac{f(r_t \mid S_t = k, \theta_k) \cdot \xi_{t|t-1}(k)}{\sum_{j=0}^{K-1} f(r_t \mid S_t = j, \theta_j) \cdot \xi_{t|t-1}(j)},
\label{eq:update}
\end{equation}
where $f(r_t \mid S_t = k, \theta_k)$ is the emission density for state $k$ with parameters $\theta_k$. The log-sum-exp trick is applied in the denominator to prevent numerical underflow: $\log \sum_j \exp(a_j) = \max_j a_j + \log \sum_j \exp(a_j - \max_j a_j)$. The Hamilton filter has $O(TK^{2})$ computational complexity, which is manageable for $K = 3$ and $T \leq 40,000$.

\subsection{AR(1)-MS-GARCH Emission Model}
\label{subsec:garch}
The return equation incorporates an AR(1) mean specification to capture the well-documented first-order autocorrelation in high-frequency FX returns:
\begin{equation}
r_t = \mu_{S_t} + \phi_{S_t} r_{t-1} + \varepsilon_t, 
\label{eq:return} \end{equation} where $\mu_{S_t}$ is the state-dependent intercept and $\phi_{S_t}$ is the state-dependent AR(1) coefficient. The residual $\varepsilon_t$ follows a state-dependent GARCH(1,1) process:
\begin{equation}
\sigma^{2}_{t,k} = \omega_k + \alpha_k \varepsilon^{2}_{t-1} + \beta_k \sigma^{2}_{t-1,k},
\label{eq:garch}
\end{equation}

where $\omega_k > 0$, $\alpha_k \geq 0$, and $\beta_k \geq 0$ are state-specific parameters and $\varepsilon_{t-1}$ denotes the innovation from the conditional mean equation. Following Haas et al. (2004), all regime-specific variance processes are updated using the same realized return innovation while each state maintains its own conditional variance dynamics. The unconditional variance within state $k$ is: \begin{equation} \sigma^{2}_{k,\infty} = \frac{\omega_k}{1 - \alpha_k - \beta_k}, \label{eq:uncond} \end{equation} which requires the covariance stationarity condition $\alpha_k + \beta_k < 1$ for all $k$.

Following \citet{haas2004}, the $K$ GARCH processes run independently in parallel, one per regime. This contrasts with the \citet{gray1996} path-dependent specification, in which the variance at time $t$ depends on the probability-weighted average of all past regime-specific variances. The \citet{haas2004} specification is adopted here because it makes the likelihood exactly tractable and reduces computational cost by a factor of 3--5x relative to the \citet{gray1996} collapsing approach.

The emission distribution is a standardised Student-$t$ with state-specific degrees of freedom $\nu_k$: 
\begin{equation} 
\log f(\varepsilon_t; \sigma_{t,k}, \nu_k) = \log\Gamma\left(\frac{\nu_k+1}{2}\right) - \log\Gamma\left(\frac{\nu_k}{2}\right) - \frac{1}{2}\log[\pi(\nu_k-2)] - \log\sigma_{t,k} - \frac{\nu_k+1}{2}\log\left[1 + \frac{(\varepsilon_t/\sigma_{t,k})^{2}}{\nu_k-2}\right].
\label{eq:student}
\end{equation}

The model employs staggered parameter bounds designed to encourage economically interpretable regime separation and mitigate label switching during estimation. The constraints enforce the economically motivated ordering Calm < Turbulent < Crisis in volatility and tail-risk characteristics.

These bounds are implemented as box constraints within the L-BFGS-B optimiser. The stagger on $\omega_k$ encourages higher baseline variance in higher-volatility regimes, while the stagger on $\nu_k$ encourages heavier tails in Crisis states, consistent with observed financial market behaviour.

\subsection{Time-Varying Transition Probabilities}
\label{subsec:tvtp}
Following \citet{filardo1994}, the transition probabilities are modelled as functions of an observable composite stress index $z_t$ through a multinomial logit specification: 
\begin{equation} 
p_{ij,t} = \frac{\exp(a_{ij} + \gamma_{ij} z_t)}{\sum_{k=0}^{K-1} \exp(a_{ik} + \gamma_{ik} z_t)}, 
\label{eq:tvtp} 
\end{equation} 
where $a_{ij}$ are baseline log-transition parameters, $\gamma_{ij}$ are TVTP sensitivity coefficients, and $z_t$ is the standardized exogenous driver. The softmax structure guarantees $\sum_j p_{ij,t}=1$ for all $i$ and $t$. The sensitivity coefficients are estimated subject to box constraints to prevent the exogenous driver from dominating the baseline transition structure and to preserve stable Markov dynamics during optimization.

The composite stress driver $z_t$ is constructed using timeframe-specific feature compositions. For the 1H Micro model,
\[
z_t^{(1H)} = 0.80 \cdot \text{comp\_meta\_1h\_z} + 0.10 \cdot \text{DOW\_Seasonality} + 0.10 \cdot \text{Hour\_Seasonality},
\]
where $\text{comp\_meta\_1h\_z}$ is the standardized microstructure stress signal
\[
\text{comp\_meta}_{1H} = \text{MicroVolSpike} \times \log(1+\text{JumpRatio}),
\]
combining intraday volatility acceleration with jump activity and subsequently standardized via a rolling 30-day z-score.

For the 4H Meso model,
\[
z_t^{(4H)} = 0.60 \cdot \text{comp\_meta\_4h\_z} + 0.25 \cdot \text{DOW\_Seasonality} + 0.15 \cdot \text{Macro\_Stress\_Index},
\]
where $\text{comp\_meta\_4h\_z}$ is a standardized meso-scale stress signal combining volatility acceleration and jump activity over a rolling 120-hour horizon. The Macro Stress Index is defined as
\[
\text{Macro\_Stress\_Index} = \text{VIX\_Z} + \text{Yield\_Z},
\]
where both components are rolling 30-day z-scores of the VIX and US 10-year Treasury yield, shifted by one period to enforce strict causality.

TVTP model selection employs the Akaike Information Criterion,
\[
\mathrm{AIC} = 2k - 2\log\hat{L},
\]
where $k$ denotes the number of free parameters and $\hat{L}$ is the maximized likelihood. The TVTP specification introduces an additional $K^2=9$ sensitivity parameters $\gamma_{ij}$ relative to the static model. The improvement criterion is defined as
\[
\Delta\mathrm{AIC} = \mathrm{AIC}_{\text{static}} - \mathrm{AIC}_{\text{TVTP}}.
\]
A $\Delta\mathrm{AIC}$ greater than 10 is generally interpreted as strong evidence in favor of the TVTP specification. The empirical results are $\Delta\mathrm{AIC}=+690.7$ (4H), $\Delta\mathrm{AIC}=+499.9$ (1H), and $\Delta\mathrm{AIC}=-2170.24$ (1D), indicating strong support for time-varying transition dynamics at the meso and micro horizons, while providing evidence against the TVTP extension at the daily frequency.

\subsection{Penalised Maximum Likelihood Estimation}
\label{subsec:mle}
The objective function minimised by the L-BFGS-B optimiser is the penalised negative log-likelihood:
\begin{align}
\mathcal{L}_{\text{pen}}(\theta) &= -\log \mathcal{L}(\theta) + \sum_{k=0}^{K-1} \lambda_k (-\log p_{kk}) \nonumber \\
&\quad + \lambda_{\text{ord}} \sum_{k=0}^{K-2} \max(0,\ p_{k+1,k+1} - p_{k,k})^{2} \nonumber \\
&\quad + \lambda_{\text{stat}} \sum_{k=0}^{K-1} \max(0,\ \alpha_k + \beta_k - 0.999)^{2},
\label{eq:penlik}
\end{align}
where $-\log \mathcal{L}(\theta)$ is the standard Hamilton filter negative log-likelihood. The second term is a log-Dirichlet stickiness prior with per-state weights $\lambda_k$ (state-specific persistence penalties are employed, with larger penalties assigned to lower-volatility states in order to encourage realistic regime persistence), which penalises low diagonal persistence. An ordering penalty is imposed to discourage violations of the persistence hierarchy p00 $\geq$ p11 $\geq$ p22. The fourth term is the GARCH stationarity penalty, which prevents explosive variance dynamics.

The L-BFGS-B optimiser is chosen for three reasons: the parameter space is high-dimensional (27 static or 36 TVTP parameters) but smooth; box constraints from the staggered bounds are natively supported; and the limited-memory Hessian approximation is computationally efficient at this parameter scale. The label-switching problem---the tendency of mixture model optimisers to permute state labels across runs---is resolved through variance-based state sorting: states are relabelled in ascending order of their empirical within-state return variance, ensuring that State 0 always corresponds to the lowest-variance (Calm) regime.

\subsection{Shannon Entropy and Regime Uncertainty}
\label{subsec:entropy}
The Shannon entropy of the filtered state probability vector provides a scalar measure of regime uncertainty:
\begin{equation}
H_t = -\sum_{k=0}^{K-1} \pi_t(k) \log \pi_t(k),
\label{eq:entropy}
\end{equation}
with maximum value $\log K = \log 3 \approx 1.099$ for $K = 3$ states. The normalised entropy is:
\begin{equation}
\tilde{H}_t = \frac{H_t}{\log K} \in [0, 1].
\label{eq:norm_entropy} \end{equation} The entropy threshold $\tilde{H}_t > 0.85$ is used to identify high-uncertainty environments, preventing trading in the most uncertain regime environments. 

%% ============================================================
\section{Data and Feature Engineering}
\label{sec:data}
%% ============================================================
The primary dataset consists of EUR/USD 1-minute mid-price (bid-ask midpoint) data spanning 2015–2025. The data are resampled to three frequencies using strictly causal aggregation: 1H (last close of each hour), 4H (last close of each 4-hour bar), and 1D (last close of each trading day). Log returns are computed as:
\begin{equation}
r_t = \log(P_t / P_{t-1}) \times 100,
\label{eq:logret}
\end{equation}
where the multiplication by 100 converts to percentage log returns, improving numerical conditioning for the GARCH parameter estimation. The train/test split date is 2021-01-01. The resulting dataset sizes are: 1D (1,871 train / 1,562 test bars), 4H (9,654 / 8,055), and 1H (37,345 / 31,152). The out-of-sample period comprises 31,152 hourly observations across 16 walk-forward analysis quarters.

The TVTP driver construction employs three layers of causal isolation to prevent look-ahead bias. First, all TVTP drivers use a one-period lag to prevent endogenous leakage. Second, the train/test split uses strict boundary slicing with a one-hour offset. Third, the maximum horizon trimming removes the last 24 training observations from the feature matrix to prevent multi-step forward target overlap.

The composite meta-signal is constructed as:
\begin{equation}
\text{composite\_meta} = \text{Vol\_Spike} \times \log(1 + \text{Jump\_Ratio}),
\label{eq:composite}
\end{equation}
where $\text{Vol\_Spike} = \sigma_{15\text{min}} / \sigma_{24\text{h}}$ is the ratio of 15-minute realised volatility to 24-hour baseline volatility, and $\text{Jump\_Ratio} = \text{Realized\_Var}_{24\text{h}} / \text{Bipower\_Var}$ is the ratio of realised variance to bipower variation (a jump-robust estimator). The product combines simultaneous volatility spikes and jump activity, the hallmark of intraday crisis events. All composite signals are standardised via rolling 30-day z-scores and clipped to $\pm 3\sigma$ to prevent outlier contamination.

The 27-expert Mixture-of-Experts feature matrix includes 47 features spanning mean reversion signals, momentum signals, volatility regime indicators, derived HMM features including entropy measures, dwell-time statistics, confidence scores, forward crisis hazard estimates, and cross-scale disagreement metrics. The top feature by absolute WFA Ridge coefficient is "Sig\_MR" with coefficient +0.001086, consistent with EUR/USD's well-documented mean-reverting behaviour at the 1H scale.

%% ============================================================
\section{Model Architecture}
\label{sec:arch}
%% ============================================================
The implementation consists of an AR(1)-MS-GARCH framework integrated within a broader machine-learning pipeline.

The static specification contains 27 optimization parameters per timeframe: 18 state-conditional emission parameters $\mu_k, \phi_k, \omega_k, \alpha_k, \beta_k, \nu_k$ for $k \in \{0, 1, 2\}$ and 9 unconstrained transition parameters. The transition matrix is mapped to a row-stochastic probability matrix through a softmax normalization. Although the resulting transition matrix has only six statistical degrees of freedom, the optimization is performed in a 9-dimensional unconstrained parameter space for numerical stability.

For the TVTP specification, an additional 9 transition-sensitivity coefficients $\gamma_{ij}$ are introduced, increasing the parameter count from 27 to 36 optimization parameters per timeframe. The Hamilton filter is implemented via Numba-JIT compiled kernels, achieving C-level execution speed. The log-sum-exp trick is applied throughout for numerical stability.

The stickiness prior vector for the 1H Micro model encodes the empirical prior that Calm regimes are highly persistent, Turbulent regimes are moderately persistent, and Crisis regimes are short-lived.

The 27-expert Mixture-of-Experts architecture constructs the joint probability tensor as:
\begin{equation}
\mathcal{P}_t(i,j,k) = \pi_t^{(1\text{D})}(i) \times \pi_t^{(4\text{H})}(j) \times \pi_t^{(1\text{H})}(k), \quad i,j,k \in \{0, 1, 2\},
\label{eq:tensor}
\end{equation}
implemented via \texttt{numpy.einsum} for computational efficiency. The resulting 27-dimensional vector is normalised to sum to 1 and serves as soft routing weights for the 27 independent RidgeCV expert models. Expert models are only trained when their state combination has at least 1,000 training observations; data-sparse combinations fall back to the global model. Forward-fill alignment propagates daily and 4H state probabilities to the hourly index using the unique causal method (forward-fill alignment), ensuring no look-ahead bias.

The Regime Disagreement metric captures cross-scale conflict:
\begin{equation}
\text{Disagreement}_t = \frac{1}{K} \sum_{k=0}^{K-1} \mathrm{std}\left(\pi_t^{(1\text{H})}(k),\ \pi_t^{(4\text{H})}(k),\ \pi_t^{(1\text{D})}(k)\right),
\label{eq:disagreement}
\end{equation}
where the standard deviation is computed across the three timescales for each state $k$. High disagreement indicates that the three timescales are assigning different regime probabilities, which is itself an informative signal about regime uncertainty and transition risk.

%% ============================================================
\section{Multi-Scale Architecture and Walk-Forward Analysis}
\label{sec:multiscale}
%% ============================================================
\subsection{The Scale Mismatch Problem}
\label{subsec:mismatch}
The fundamental motivation for the multi-scale architecture is the incompatibility of regime transition rates across timescales. At the daily frequency, macro regime transitions are rare: $P[\text{Calm} \to \text{Crisis}] \approx 0.001$ per day, corresponding to a direct jump probability of 0.0215 in the estimated 1D transition matrix. At the hourly frequency, microstructure transitions are far more frequent: $P[\text{Calm} \to \text{Turbulent}] \approx 0.08$ per hour, as reflected in the 1H transition matrix entry $p_{01} = 0.0824$. A single $K \times K$ transition matrix cannot simultaneously encode both rates; the optimiser will find a compromise that satisfies neither constraint well.

Four independent arguments support the multi-scale architecture. The \textit{computational argument} notes that a single HMM on 1H data with $K = 9$ states (to capture $3 \times 3$ macro-micro combinations) would have $O(TK^{2}) = O(50,000 \times 81)$ complexity per likelihood evaluation, compared to three separate $K = 3$ models with $O(50,000 \times 9)$ each. The \textit{statistical argument} notes that a single model must simultaneously capture dynamics at incompatible timescales, producing a misspecified transition matrix. The \textit{interpretability argument} notes that the multi-scale architecture produces separately interpretable outputs at each scale. The empirical argument notes that the three models produce materially different regime allocations, persistence characteristics, and transition dynamics, indicating that each timescale captures distinct aspects of market behaviour.

\subsection{Three-Layer Information Hierarchy}
\label{subsec:hierarchy}
The three models capture distinct layers of market dynamics. The 1D Macro model captures macro-structural regime changes driven by central bank policy cycles, geopolitical risk episodes, and long-horizon capital flows. The 4H Meso model captures medium-term flow regimes driven by institutional positioning cycles, options expiry dynamics, and multi-day momentum. The 1H Micro model captures intraday microstructure regimes driven by session transitions, news flow, and short-term liquidity dynamics.

\begin{table}[htbp]
\centering
\caption{Three-Layer Architecture: Comparative Summary}
\label{tab:architecture}
\begin{tabular}{lccc}
\toprule
Dimension & 1D Macro & 4H Meso & 1H Micro \\
\midrule
Training bars & 1,871 & 9,654 & 37,345 \\
Test bars & 1,562 & 8,055 & 31,152 \\
TVTP specification & Static & $\Delta\mathrm{AIC}=+690.7$ & $\Delta\mathrm{AIC}=+499.9$ \\
AIC & $+2170.24$ & $-6880.72$ & $-80160.25$ \\
Calm dwell time & 45.1 days & 10.8 days & 10.7 hours \\
Turbulent dwell time & 13.9 days & 1.9 days & 5.1 hours \\
Crisis dwell time & 7.6 days & 0.6 days & 3.9 hours \\
OOS Calm allocation & 82.3\% & 8.4\% & 65.6\% \\
OOS Turbulent allocation & 10.1\% & 71.6\% & 32.5\% \\
OOS Crisis allocation & 7.7\% & 20.0\% & 2.0\% \\
Brier Skill Score & $+0.2132$ & $+0.0874$ & $+0.1185$ \\
RCM Clarity & 27.6\% & 52.4\% & 54.0\% \\
\bottomrule
\end{tabular}
\end{table}

The 4H Meso model's Turbulent-dominant allocation (71.6\%) is expected for EUR/USD during the 2021--2025 period, which was characterised by persistent institutional flow activity, elevated options market activity, and frequent ECB/Fed communication events. The 1D Macro model's Calm-dominant allocation (82.3\%) reflects the daily timescale's lower sensitivity to intraday noise; at the daily level, EUR/USD spends most of its time in a low-volatility, mean-reverting regime.

\subsection{Rolling Walk-Forward Analysis}
\label{subsec:wfa}
The walk-forward analysis protocol employs 16 non-overlapping quarters from 2021-Q1 through 2024-Q4. At each quarter boundary, the three HMM models are re-estimated on a rolling 5-year window of approximately 31,000 hourly bars. The design fuses the training tail to the test head, ensuring that rolling features (such as 24-hour moving averages) roll seamlessly across quarter boundaries without introducing NaN values or look-ahead bias.

The warm start strategy initialises the L-BFGS-B optimiser at the parameter estimates from the previous quarter, rather than at the default initial guess. This provides three benefits: reduced optimisation time (starting close to the solution requires fewer iterations); improved label consistency (warm start anchors the solution near the previous quarter's labelling, reducing label switching); and smooth tracking of gradual parameter drift (regime characteristics evolve slowly over time, and warm start allows the model to track this drift quarter-by-quarter).

The 16 quarters span a remarkably diverse set of macro environments: post-COVID recovery with EUR/USD ranging between 1.17 and 1.23 (2021); the Fed tightening cycle with EUR/USD declining to parity and Russia-Ukraine shock (2022); banking stress with SVB and Credit Suisse failures (2023); and carry trade unwind and normalisation (2024). This diversity provides a rigorous test of the model's robustness across different market regimes.

%% ============================================================
\section{Statistical Validation}
\label{sec:validation}
%% ============================================================
\subsection{Transition Matrix Analysis}
\label{subsec:trans}
Under the first-order Markov property, the expected dwell time in state $k$ follows from the geometric distribution:
\begin{equation}
\E[T_k] = \frac{1}{1 - p_{kk}}.
\label{eq:dwell}
\end{equation}
The three estimated transition matrices are as follows. For the 1D Macro model (static):
\begin{equation}
P^{(1\text{D})} = \begin{pmatrix} 0.9778 & 0.0007 & 0.0215 \\ 0.0025 & 0.9283 & 0.0693 \\ 0.0768 & 0.0548 & 0.8685 \end{pmatrix},
\label{eq:P1d}
\end{equation}
yielding dwell times of 45.1 days (Calm), 13.9 days (Turbulent), and 7.6 days (Crisis). For the 4H Meso model (TVTP):
\begin{equation}
P^{(4\text{H})} = \begin{pmatrix} 0.9846 & 0.0151 & 0.0003 \\ 0.0136 & 0.9125 & 0.0738 \\ 0.0177 & 0.2541 & 0.7282 \end{pmatrix},
\label{eq:P4h}
\end{equation}
yielding dwell times of 10.8 days (Calm), 1.9 days (Turbulent), and 0.6 days (Crisis). For the 1H Micro model (TVTP):
\begin{equation}
P^{(1\text{H})} = \begin{pmatrix} 0.9062 & 0.0824 & 0.0113 \\ 0.1437 & 0.8028 & 0.0535 \\ 0.0332 & 0.2228 & 0.7440 \end{pmatrix},
\label{eq:P1h}
\end{equation}
yielding dwell times of 10.7 hours (Calm), 5.1 hours (Turbulent), and 3.9 hours (Crisis). The hierarchical structure is immediately apparent: macro regimes are slow-moving structural states, meso regimes are medium-frequency cyclical states, and micro regimes are high-frequency transient states.

\subsection{Regime Classification Measure}
\label{subsec:rcm}
The Regime Classification Measure (RCM) quantifies the sharpness of regime assignments:
\begin{equation}
\mathrm{RCM} = 100 \times \left[1 - \frac{K}{K-1} \cdot \frac{1}{T} \sum_{t=1}^T \sum_{k=0}^{K-1} \pi_t(k)(1 - \pi_t(k))\right].
\label{eq:rcm}
\end{equation}
RCM $= 100\%$ indicates perfect classification (all probabilities are 0 or 1); RCM $= 0\%$ indicates uniform uncertainty. The empirical results are 27.6\% (1D), 52.4\% (4H), and 54.0\% (1H). The values indicate moderate regime classification confidence, which is expected in FX markets where emission distributions overlap more than in equity markets.

\subsection{Brier Skill Scores}
\label{subsec:bss}
The Brier Skill Score \citep{brier1950} measures probabilistic forecast skill relative to a climatological baseline:
\begin{equation}
\mathrm{BSS} = 1 - \frac{\mathrm{BS}_{\text{model}}}{\mathrm{BS}_{\text{ref}}} = 1 - \frac{(1/T)\sum_t (p_t - y_t)^2}{\bar{p}(1-\bar{p})}, \label{eq:bss} \end{equation} where $p_t = \pi_t(\text{Crisis})$ is the model's crisis probability forecast and $y_t = 1[|r_t| > \text{85th percentile}]$ is the realised high-volatility indicator. The empirical results are +0.2132 (1D), +0.0874 (4H), and +0.1185 (1H). All three timeframes achieve positive BSS, confirming that the MS-GARCH model has genuine skill in predicting high-volatility periods beyond the climatological base rate.

\subsection{ARCH-LM Residual Diagnostics}
\label{subsec:arch_test}
The ARCH-LM test \citep{engle1982} applied to the standardised residuals $\hat{\varepsilon}_t = r_t / \sigma_{t,\hat{S}_t}$ reveals persistent conditional heteroskedasticity in all three timeframes: $\mathrm{LM} = 12.19$ ($p = 0.032$) for 1D, $\mathrm{LM} = 16.75$ ($p = 0.005$) for 4H, and $\mathrm{LM} = 38.49$ ($p = 0.000$) for 1H. The residual ARCH is most severe at the 1H Micro level, which is expected given that intraday volatility clustering is stronger than daily clustering. The unconditional ACF of squared returns is 0.2579, reduced to a within-state maximum of 0.1889 by the MS-GARCH model, representing a 26.8\% absorption of total volatility clustering. The residual ARCH is an acknowledged limitation addressed in Section~\ref{sec:limits}.

\subsection{TVTP Value Test}
\label{subsec:tvtp_test}
The TVTP value test \citep{akaike1974} compares the AIC of the TVTP model to the static model: $\Delta\mathrm{AIC} = \mathrm{AIC}_{\text{static}} - \mathrm{AIC}_{\text{TVTP}}$. The empirical results confirm strong TVTP superiority at the 4H and 1H scales ($\Delta\mathrm{AIC} = +690.7$ and $+499.9$, far exceeding the threshold of 10 for decisive evidence), while the 1D model correctly employs a static specification (positive AIC $= +2170.24$ confirms near-overparameterisation).

\subsection{VaR Breach Rate}
\label{subsec:var}
The 99\% VaR breach rate is 0.54\% versus the target of 1.00\%, indicating that the model is conservative in its tail risk estimation. This is a desirable property for risk management applications: the model overestimates tail risk, providing a safety margin against unexpected extreme events.

\subsection{Consolidated Validation Scorecard}
\label{subsec:scorecard}
\begin{table}[htbp]
\centering
\caption{Consolidated Statistical Validation Scorecard}
\label{tab:scorecard}
\begin{tabular}{lccc}
\toprule
Test & 1D Macro & 4H Meso & 1H Micro \\
\midrule
Ordering constraint & Valid & Valid & Valid \\
RCM Clarity & 27.6\% & 52.4\% & 54.0\%  \\
Brier Skill Score & $+0.2132$ & $+0.0874$ & $+0.1185$ \\
ARCH-LM $p$-value & 0.032 & 0.005 & 0.000 \\
TVTP $\Delta$AIC & N/A (static) & $+690.7$ & $+499.9$ \\
VaR 99\% breach rate & 0.54\% & --- & --- \\
Label integrity (OOS) & Monotonic & Monotonic & Monotonic \\
\bottomrule
\end{tabular}
\end{table}

%% ============================================================
\section{Volatility Forecasting Benchmark}
\label{sec:benchmark}
%% ============================================================
The GARCH(1,1) benchmark \citep{bollerslev1986} is estimated on the training set using maximum likelihood, yielding parameters $\hat{\omega} = 0.001680$, $\hat{\alpha} = 0.3038$, $\hat{\beta} = 0.6076$, and persistence $\hat{\alpha} + \hat{\beta} = 0.9114$. The anomalously high $\hat{\alpha} = 0.3038$---three to six times above the typical range of 0.05--0.10 for EUR/USD---is a diagnostic signature of model misspecification. The single-regime GARCH model, confronted with a sample containing multiple structural regimes, absorbs the regime-switching dynamics into its shock-response parameter. The relatively low $\hat{\beta} = 0.6076$ (compared to typical values of 0.85--0.92) is the counterpart: the combination of high $\hat{\alpha}$ and low $\hat{\beta}$ creates a model that reacts aggressively to individual shocks but does not sustain elevated volatility for long, mimicking regime-switching behaviour through a single-state mechanism.

Both the MS-GARCH and GARCH forecasts are smoothed using an exponentially weighted moving average with span 24 before comparison, to reduce short-term forecast noise and facilitate comparison of forecast dynamics.

\begin{table}[htbp]
\centering
\caption{Volatility Forecast Quality: MS-GARCH vs.\ Causal GARCH(1,1)}
\label{tab:benchmark}
\begin{tabular}{lccc}
\toprule
Metric & GARCH(1,1) & MS-GARCH & Improvement \\
\midrule
RMSE (smoothed) & 0.037411 & 0.036711 & $-1.87\%$ \\
Spearman IC (smoothed) & 0.5264 & 0.5371 & $+1.07$ pp \\
DM statistic & \multicolumn{2}{c}{$+4.7040$} & --- \\
DM $p$-value & \multicolumn{2}{c}{$1.28 \times 10^{-6}$} & --- \\
Error mean & 0.002324 & 0.001256 & $-46\%$ \\
Error std & 0.037339 & 0.036690 & $-1.74\%$ \\
\bottomrule
\end{tabular}
\end{table}

\begin{figure}[htbp]
    \centering
    \includegraphics[width=0.8\linewidth]{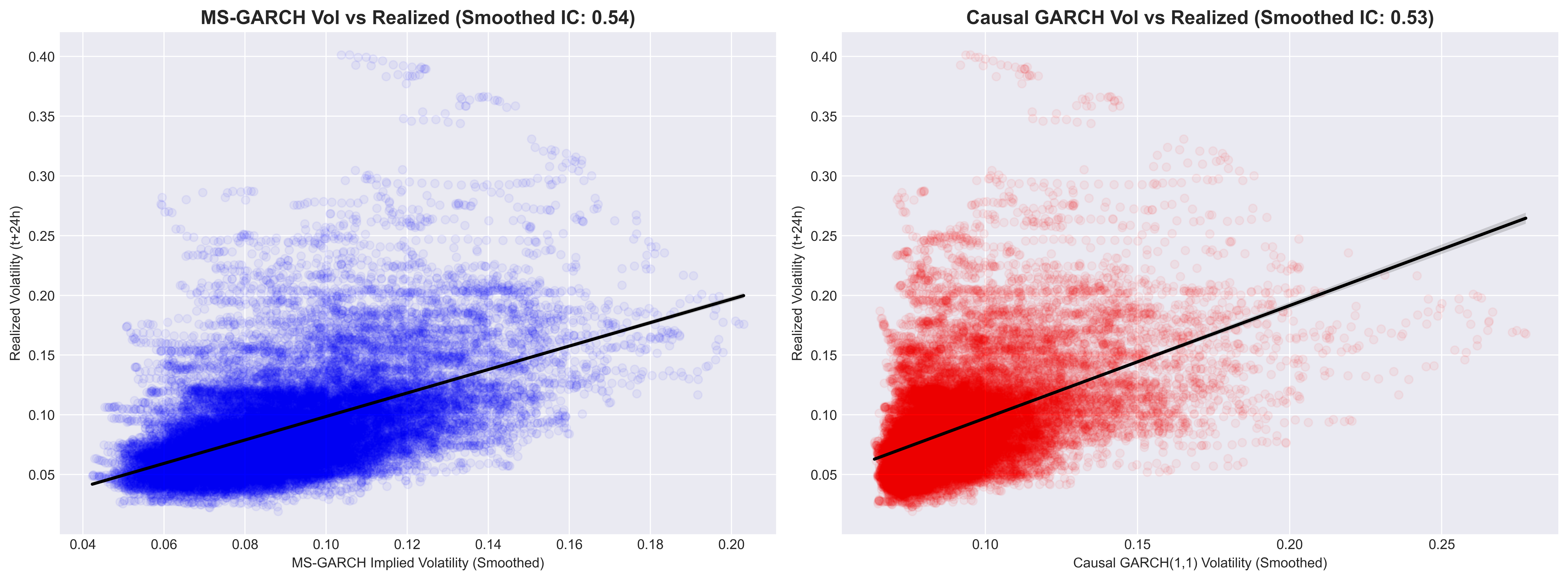}
    \caption{Realized vs. Implied Volatility Forecasts (Smoothed MS-GARCH vs. Causal GARCH)}
    \label{fig:volforecast}
\end{figure}

The Diebold--Mariano test \citep{diebold1995} formally confirms MS-GARCH superiority: $\mathrm{DM} = +4.7040$, $p = 1.28 \times 10^{-6}$, rejecting the null hypothesis of equal predictive accuracy at any conventional significance level. The higher smoothed IC of the MS-GARCH model reflects superior stability and regime-conditioned forecasting performance.

%% ============================================================
\section{Empirical Results}
\label{sec:results}
%% ============================================================
\subsection{Out-of-Sample Regime Allocations}
\label{subsec:alloc}
The OOS regime allocations reflect the empirical distribution of EUR/USD microstructure states over the 2021--2025 period. At the 1D Macro scale, the model allocates 82.3\% to Calm, 10.1\% to Turbulent, and 7.7\% to Crisis. At the 4H Meso scale, the allocation is 8.4\% Calm, 71.6\% Turbulent, and 20.0\% Crisis. At the 1H Micro scale, the allocation is 65.6\% Calm, 32.5\% Turbulent, and 2.0\% Crisis.

The striking divergence between timeframes is economically meaningful. The 1D model identifies the 2021--2025 period as predominantly calm at the macro level, consistent with the daily timescale's lower sensitivity to intraday noise. The 4H model identifies persistent turbulence (71.6\%), reflecting the sustained institutional flow activity, elevated options market activity, and frequent ECB/Fed communication events that characterised this period. The 1H model's 2.0\% Crisis allocation reflects the rarity of genuine acute hourly stress episodes.

The Rolling Walk-Forward (WFA) quarterly analysis reveals substantial time-variation in regime allocations. In 2021-Q3 (July--September 2021), the 1D Macro model allocated 100\% to Calm---a period of EUR/USD stability following the post-COVID recovery. In contrast, 2022-Q3 (July--September 2022) saw 77\% Crisis allocation at the Macro level, corresponding to the peak of the Fed tightening cycle and EUR/USD falling below parity. This time-variation confirms that the model is capturing genuine regime dynamics rather than fitting a stationary mixture.

\subsection{Conditional Return Profiles}
\label{subsec:profiles}
\begin{table}[htbp]
\centering
\caption{Conditional Return Profiles by Regime and Timescale (OOS: 2021--2025)}
\label{tab:profiles}
\begin{tabular}{llccc}
\toprule
Timeframe & Regime & Ann.\ Vol & Fisher Kurtosis & OOS Alloc. \\
\midrule
1D Macro & Calm & 5.72\% & 1.0 & 82.3\% \\
1D Macro & Turbulent & 7.88\% & 4.4 & 10.1\% \\
1D Macro & Crisis & 13.45\% & -0.7 & 7.7\% \\
\midrule
4H Meso & Calm & 2.19\% & 0.2 & 8.4\% \\
4H Meso & Turbulent & 5.41\% & 3.0 & 71.6\% \\
4H Meso & Crisis & 13.22\% & 2.2 & 20.0\% \\
\midrule
1H Micro & Calm & 3.70\% & 1.4 & 65.6\% \\
1H Micro & Turbulent & 10.92\% & 2.2 & 32.5\% \\
1H Micro & Crisis & 24.72\% & 3.2 & 2.0\% \\
\bottomrule
\end{tabular}
\end{table}

Some patterns emerge immediately from Table~\ref{tab:profiles}. First, the volatility ordering Calm $<$ Turbulent $<$ Crisis holds monotonically across all three timescales, confirming that no label switching has occurred. This is validated by Test 1 (OOS Volatility Stratification), which reports annualised volatilities of 3.70\%, 10.92\%, and 24.72\% for the 1H Micro states.

\begin{figure}[htbp]
    \centering
    \includegraphics[width=0.8\linewidth]{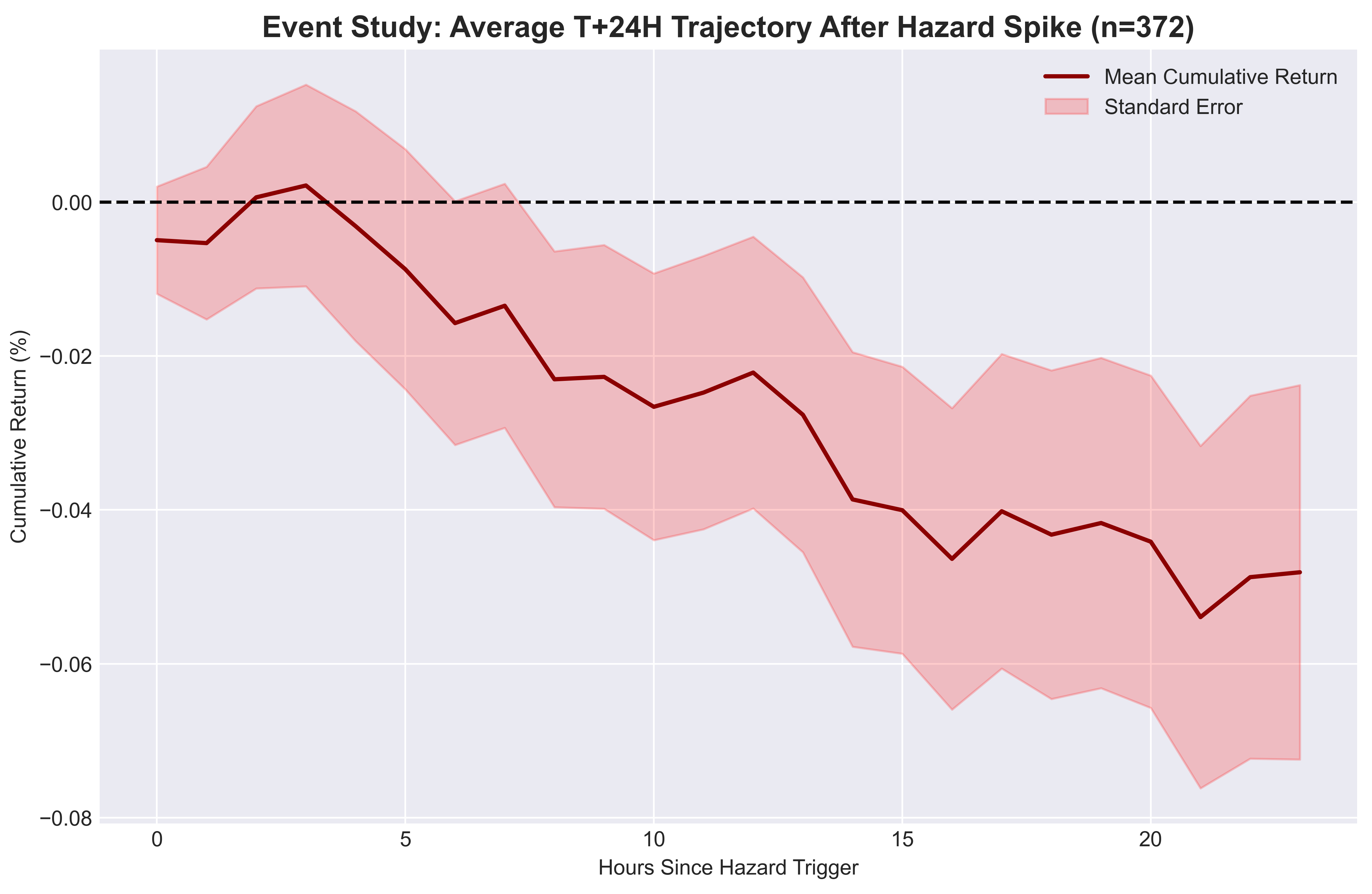}
    \caption{Average T+24H Trajectory After Hazard Spike}
    \label{fig:forwardhazard}
\end{figure}

\subsection{OOS Regime Validation Diagnostics}
\label{subsec:tests}
\begin{table}[htbp]
\centering
\caption{Out-of-Sample Regime Validation Results (Strategy-Independent, OOS 2021--2025)}
\label{tests}
\small
\setlength{\tabcolsep}{5pt} % Tightens horizontal column padding to fit perfectly inside margins
\begin{tabular}{lll}
\toprule
Test & Result & Statistic\\
\midrule
Test 1: Volatility Stratification & Calm $<$ Turb $<$ Crisis & Monotonic \\
Test 2: CVaR-99 Monotonicity & -32.3 $<$ -49.9 $<$ -55.9 bps & Monotonic \\
Test 3: OOS Churn Stability & 4,365 flips / 24,955 hours & 5.72h avg dwell \\
Test 4: KS Distributional Purity & Calm vs.\ Turb & $p = 1.35 \times 10^{-153}$ \\
Test 4: KS Distributional Purity & Turb vs.\ Crisis & $p = 1.03 \times 10^{-38}$ \\
Test 7: ACF Absorption & 0.2579 $\to$ 0.1889 & 26.8\% reduction \\
Test 8: Transition Stability & Empirical Calm persistence 0.876 & --- \\
\bottomrule
\end{tabular}
\end{table}

\begin{figure}[htbp]
    \centering
    \makebox[\linewidth][c]{%
        \includegraphics[width=0.6\linewidth, height=10cm, keepaspectratio]{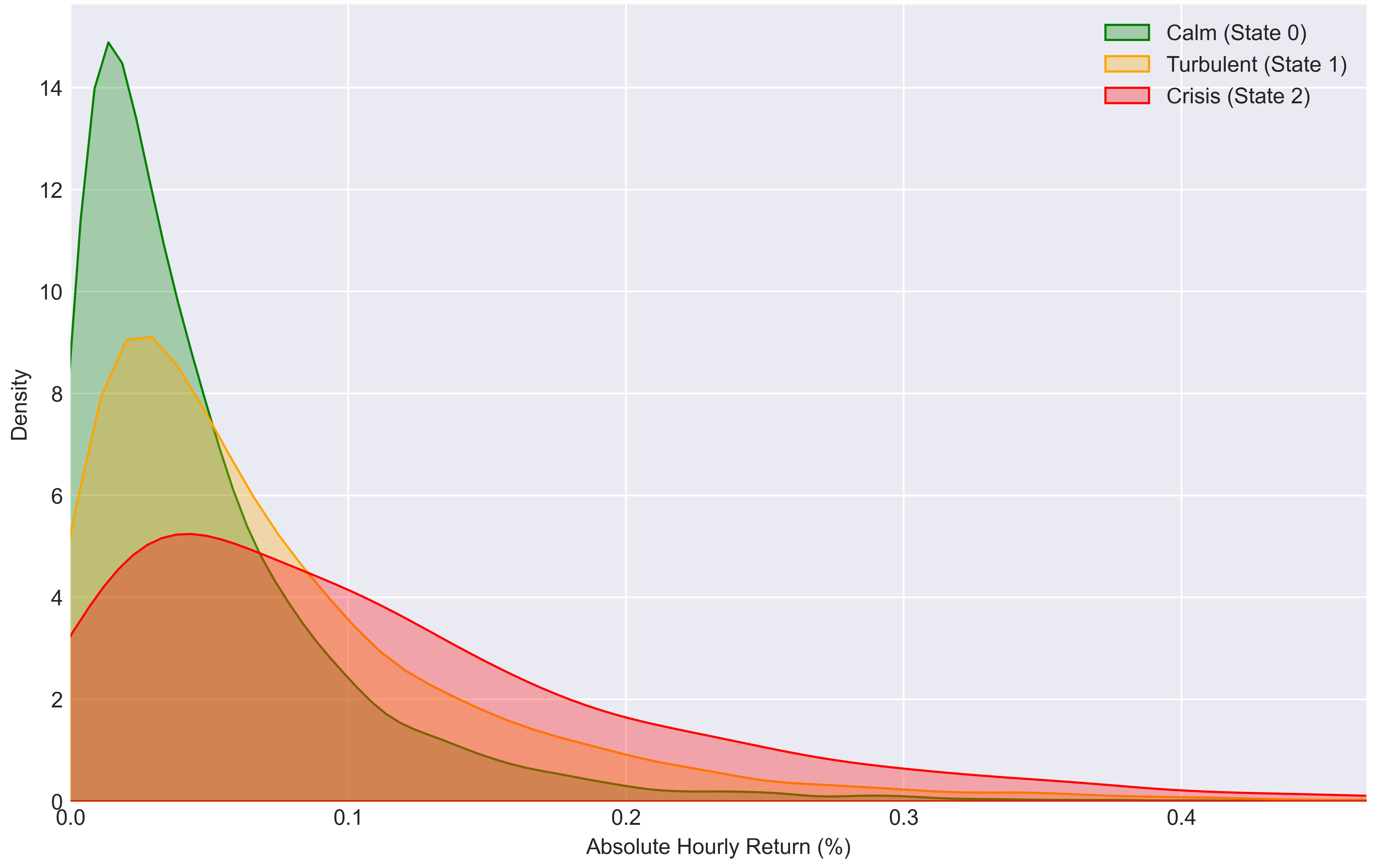}%
    }
    % Main short title for the List of Figures
    \caption{Out-of-Sample Return Distribution (1H) Separation Across Hidden States.}
    \label{fig:regimeseparation}
    
    \vspace{5pt}
    
    % Multi-line technical explanation block placed directly underneath
    \vspace{4pt}
    \begin{minipage}{0.8\linewidth} % Matches or slightly exceeds your image width for clean text alignment
        \footnotesize
         Kernel density estimates of absolute hourly returns for Calm, Turbulent, and Crisis regimes. The distributions progressively shift toward larger return magnitudes, supporting the volatility ordering observed across all temporal layers.
    \end{minipage}
\end{figure}

The KS test results provide strong evidence that the three-state model has identified materially distinct market regimes, not arbitrary partitions of a homogeneous distribution.

\subsection{Three-Dimensional Volatility Tensor}
\label{subsec:tensordim}
To facilitate interpretation of the 27 joint state combinations, we group them into a small number of economically meaningful meta-regimes based on recurring cross-scale alignment patterns.

\begin{table}[htbp]
\centering
\caption{Meta-Regime Attribution: WFA Rolling Results (2021--2025)}
\label{tab:metaregime}
\small
\setlength{\tabcolsep}{6pt} % Optimizes horizontal spacing
\begin{tabular}{lp{5.8cm}cc} % Fixed width for column 2 forces neat wrapping inside margins
\toprule
Meta-Regime Cluster & Dominant Tensor Configurations & Time Active & Mean IC \\
\midrule
Synchronized Turbulence & 1D:Turb $\cdot$ 4H:Turb $\cdot$ 1H:Turb & 45.09\% & +0.0312 \\
\addlinespace
Transitional Regime & Mixed Calm/Turb/Crisis states without full alignment & 28.69\% & +0.0330 \\
\addlinespace
Localized Micro Shock & 1D:Calm $\cdot$ 4H:Calm $\cdot$ 1H:Crisis & 13.55\% & +0.0075 \\
\addlinespace
Macro Stress Divergence & 1D:Crisis $\cdot$ 4H:Crisis $\cdot$ 1H:Calm & 5.94\% & +0.0544 \\
\addlinespace
Total Calm & 1D:Calm $\cdot$ 4H:Calm $\cdot$ 1H:Calm & 4.01\% & $-$0.0226 \\
\addlinespace
Total Crisis & 1D:Crisis $\cdot$ 4H:Crisis $\cdot$ 1H:Crisis & 2.73\% & $-$0.0436 \\
\bottomrule
\end{tabular}
\end{table}

The Macro Stress Divergence meta-regime (5.94\% frequency) produces the highest IC (+0.0544), suggesting that cross-scale disagreement---macro and meso stress with micro calm---is associated with stronger predictive relationships. The Total Crisis meta-regime (2.73\%) produces negative IC (-0.0436), suggesting that fully synchronised crisis conditions exhibit different predictive characteristics from cross-scale disagreement states.

\subsection{Shannon Entropy Filter}
\label{subsec:entropy_results}
The Shannon entropy measure provides a quantitative view of regime uncertainty. Lower entropy values correspond to more concentrated state probabilities and greater confidence in regime assignment, whereas higher entropy values indicate greater uncertainty and overlap between competing regime classifications. The quarterly entropy variation ranges from a minimum of 0.408 (2021-Q3, post-COVID calm with clear regime signals) to a maximum of 0.774 (2022-Q4, Fed pivot uncertainty creating ambiguous regime signals).

\subsection{Formal Hypothesis Tests}
\label{subsec:hypotheses}
\begin{table}[htbp]
\centering
\caption{Formal Hypothesis Test Results}
\label{tab:hypotheses}
\begin{tabular}{llccc}
\toprule
Hypothesis & Test & Statistic & $p$-value \\
\midrule
H1: MS-GARCH $>$ GARCH & Diebold--Mariano & $\mathrm{DM} = +4.7040$ & $1.28 \times 10^{-6}$ \\
H2: IC $> 0.015$, $p < 0.05$ & IC significance test & IC $= +0.0252$ & $6.75 \times 10^{-5}$ \\
\bottomrule
\end{tabular}
\end{table}

\begin{figure}[htbp]
    \centering
    \includegraphics[width=0.8\linewidth]{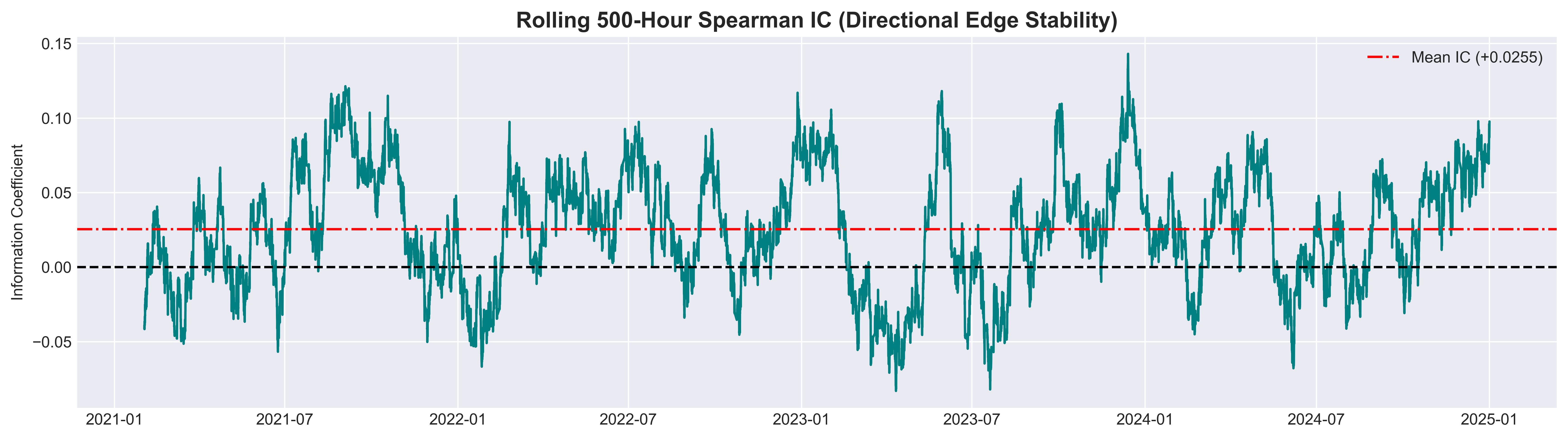}
    \caption{Rolling 500-Hour Spearman Information Coefficient (Directional Edge Stability)}
    \label{fig:icstability}
\end{figure}

%% ============================================================
\section{Limitations and Future Research}
\label{sec:limits}
%% ============================================================
\subsection{Residual ARCH Effects}
\label{subsec:arch_limit}
The ARCH-LM test \citep{engle1982} reveals statistically significant residual heteroskedasticity in all three timeframes ($\mathrm{LM} = 12.19$, $p = 0.032$ for 1D; $\mathrm{LM} = 16.75$, $p = 0.005$ for 4H; $\mathrm{LM} = 38.49$, $p = 0.000$ for 1H). The fundamental cause is the path-independence assumption of the \citet{haas2004} specification: when the market transitions from Crisis back to Calm, the Calm GARCH variance immediately reverts to its unconditional level, ignoring the elevated volatility inherited from the Crisis state. This creates a discontinuity in the implied volatility path that manifests as residual ARCH in the standardised residuals.

The \citet{gray1996} path-dependent GARCH formulation addresses this by making the variance a function of the full history of filtered probabilities: $\sigma^{2}_{t,k} = \omega_k + \alpha_k \varepsilon^{2}_{t-1} + \beta_k \E[\sigma^{2}_{t-1} \mid r_{t-2}, \ldots, r_{1}]$, where $\E[\sigma^{2}_{t-1}] = \sum_j \pi_{t-1}(j) \sigma^{2}_{t-1,j}$. This ``collapsing'' approach ensures continuity across regime transitions but increases computational cost by a factor of 3--5.

\subsection{RCM Value}
\label{subsec:rcm_limit}
RCM values of 27.6\% (1D), 52.4\% (4H), and 54.0\% (1H). These indicate only moderate regime classification certainty. This reflects a fundamental challenge in FX regime modelling: the emission distributions of adjacent states overlap substantially. Unlike equity markets, where Calm and Crisis states have clearly separated return distributions driven by the leverage effect and volatility feedback, FX markets exhibit more continuous volatility dynamics. The Normal Inverse Gaussian distribution \citep{barndorff1997} and the Variance Gamma distribution \citep{madan1998} offer heavier tails and greater flexibility than the skew Student-$t$, potentially improving regime separation by better fitting the extreme observations that define Crisis regimes.

\subsection{Batch MLE: No Real-Time Adaptation}
\label{subsec:batch_limit}
The current L-BFGS-B batch MLE approach re-estimates all parameters at each WFA quarter boundary, creating a computational bottleneck and preventing real-time parameter updating. Sequential Monte Carlo (particle filtering) methods would enable online parameter estimation, allowing the model to adapt continuously to changing market conditions. The particle filter maintains a particle approximation to the joint posterior $P(\theta, S_t \mid r_{1}, \ldots, r_t)$ and can detect structural breaks in real time through the effective sample size (ESS) criterion.

\subsection{Univariate Framework}
\label{subsec:univariate_limit}
The current framework models EUR/USD in isolation. FX correlations spike during crisis regimes---a feature that the univariate MS-GARCH model cannot capture. A multivariate extension using regime-switching copulas would capture cross-asset contagion dynamics, providing richer signals for the TVTP drivers and more robust crisis detection. The joint emission density under a regime-switching copula is $f(r_t \mid S_t = k) = c_k(F_{1,k}(r_{1,t}), \ldots, F_{n,k}(r_{n,t})) \cdot \prod_i f_{i,k}(r_{i,t})$, where $c_k$ is the state-conditional copula.

\subsection{Markov Assumption and Duration Dependence}
\label{subsec:markov_limit}
The first-order Markov assumption implies that regime transitions depend only on the current state, not on how long the system has been in that state. \citet{durland1994} documented empirical evidence of duration dependence in US GNP growth regimes, motivating semi-Markov models in which transition probabilities depend on regime age. For EUR/USD, there is economic reason to expect duration dependence: a Calm regime that has persisted for 60 days may be more likely to transition than one that has persisted for only 5 days, as the longer-duration calm may reflect a structural equilibrium that is more resistant to perturbation.

\subsection{Single Currency Pair}
\label{subsec:single_limit}
The framework is validated exclusively on EUR/USD. Generalisation to other major pairs (GBP/USD, USD/JPY, USD/CHF) and emerging market currencies would test the robustness of the triple-timeframe architecture and the universality of the identified regime structure. Different currency pairs may require different stickiness priors, TVTP driver compositions, and parameter bounds.

%% ============================================================
\section{Conclusion}
\label{sec:conclusion}
%% ============================================================
This paper has introduced a triple-timeframe Markov-Switching GARCH framework with time-varying transition probabilities for volatility regime detection in EUR/USD. The framework makes three core contributions: a multi-scale architecture that resolves the fundamental scale mismatch problem of single-timeframe HMMs; TVTP driven by composite microstructure stress indices with decisive AIC improvements of +690.7 (4H) and +499.9 (1H) \citep{akaike1974}; and a 27-dimensional joint probability tensor enabling cross-scale interaction analysis through a Mixture-of-Experts routing architecture.

The empirical results provide evidence that the proposed framework improves volatility forecasting accuracy and produces economically interpretable regime structures. The Diebold--Mariano test \citep{diebold1995} confirms statistically superior volatility forecasting: $\mathrm{DM} = +4.7040$ ($p = 1.28 \times 10^{-6}$). The KS distributional purity test confirms that the three regimes represent genuinely distinct data-generating processes: $p = 1.35 \times 10^{-153}$ for Calm vs.\ Turbulent. The master directional IC of +0.0252 ($p = 6.75 \times 10^{-5}$) confirms statistically significant directional predictability.

The empirical results support the central thesis that volatility dynamics in EUR/USD are inherently multi-scale and regime-dependent. The proposed framework produces statistically distinct volatility states, improving volatility forecasting accuracy relative to a single-regime GARCH benchmark, and captures meaningful interactions across temporal layers through the joint probability tensor. These findings suggest that modelling volatility as a hierarchy of interacting regimes provides a richer representation of market dynamics than conventional single-timescale approaches.

The path forward is clear. NIG or VG emissions \citep{barndorff1997, madan1998} would improve regime separation by better fitting the extreme observations that define Crisis regimes. Particle filtering would enable online parameter estimation and real-time structural break detection. Regime-switching copulas would extend the framework to multivariate settings, capturing cross-asset contagion dynamics. Semi-Markov models \citep{durland1994} would address the duration dependence limitation of the first-order Markov assumption. Each of these extensions addresses a specific, quantified limitation of the current framework, providing a clear research agenda for future work.

\appendix
\newpage
%% ============================================================
\section{Key Model Parameters and Diagnostic Statistics}
\label{app:params}
%% ============================================================
\begin{table}[htbp]
\centering
\caption{Summary of Key Model Parameters and Diagnostic Statistics (OOS: 2021--2025)}
\label{tab:A1}
\begin{tabular}{lccc}
\toprule
Diagnostic / Statistic & 1D Macro & 4H Meso & 1H Micro \\
\midrule
AIC & $+2170.24$ & $-6880.72$ & $-80160.25$ \\
TVTP Justified? & No (static) & Yes ($\Delta\mathrm{AIC}=+690.7$) & Yes ($\Delta\mathrm{AIC}=+499.9$) \\
Calm Dwell Time & 45.1 days & 10.8 days & 10.7 hours \\
Turbulent Dwell Time & 13.9 days & 1.9 days & 5.1 hours \\
Crisis Dwell Time & 7.6 days & 0.6 days & 3.9 hours \\
Brier Skill Score & $+0.2132$ & $+0.0874$ & $+0.1185$ \\
RCM Clarity & 27.6\% & 52.4\% & 54.0\% \\
ARCH-LM $p$-value & 0.0323 & 0.0050 & 0.0000 \\
Calm$\to$Calm Persistence & 0.9778 & 0.9846 & 0.9062 \\
Crisis$\to$Crisis Persistence & 0.8685 & 0.7282 & 0.7440 \\
OOS Calm Allocation & 82.3\% & 8.4\% & 65.6\% \\
OOS Turbulent Allocation & 10.1\% & 71.6\% & 32.5\% \\
OOS Crisis Allocation & 7.7\% & 20.0\% & 2.0\% \\
\bottomrule
\end{tabular}
\end{table}

%% ============================================================
\section{Formal Hypothesis Test Results}
\label{app:hypotheses}
%% ============================================================
\begin{table}[htbp]
\centering
\caption{Formal Hypothesis Test Results (Phase 8)}
\label{tab:A2}
\begin{tabular}{llcc}
\toprule
Hypothesis & Test & Statistic & $p$-value \\
\midrule
H1: MS-GARCH $>$ GARCH & Diebold--Mariano & $\mathrm{DM} = +4.7040$ & $1.28 \times 10^{-6}$ \\
H2: IC $> 0.015$, $p < 0.05$ & IC significance test & IC $= +0.0252$ & $6.75 \times 10^{-5}$ \\
\bottomrule
\end{tabular}
\end{table}

\newpage

%% ============================================================
\section{WFA Quarterly Scorecard}
\label{app:wfa}
%% ============================================================
\begin{table}[htbp]
\centering
\small
\setlength{\tabcolsep}{4pt}
\caption{Selected WFA Quarterly Results (2021-Q1 through 2024-Q4)}
\label{tab:A4}
\begin{tabular}{cccccc}
\toprule
Quarter & IC & \parbox{4.2cm}{\centering Macro Alloc.\\(Calm/Turb./Crisis)} & \parbox{2cm}{\centering Micro RCM} & \parbox{2.2cm}{\centering Avg Entropy} & \parbox{2.5cm}{\centering Avg Confidence} \\
\midrule
2021-Q1 & -0.0052 & 4\%/39\%/57\% & 47\% & 0.593 & 0.743 \\
2021-Q2 & +0.0105 & 29\%/71\%/0\% & 53\% & 0.530 & 0.775 \\
2021-Q3 & +0.0699 & 100\%/0\%/0\% & 66\% & 0.408 & 0.841 \\
2021-Q4 & +0.0126 & 19\%/80\%/1\% & 61\% & 0.449 & 0.818 \\
2022-Q1 & +0.0183 & 38\%/0\%/62\% & 46\% & 0.588 & 0.728 \\
2022-Q2 & +0.0531 & 23\%/1\%/76\% & 37\% & 0.687 & 0.684 \\
2022-Q3 & +0.0284 & 6\%/16\%/77\% & 38\% & 0.696 & 0.699 \\
2022-Q4 & +0.0351 & 24\%/61\%/15\% & 29\% & 0.774 & 0.637 \\
2023-Q1 & -0.0027 & 38\%/60\%/3\% & 41\% & 0.639 & 0.706 \\
2023-Q2 & +0.0141 & 88\%/12\%/0\% & 47\% & 0.605 & 0.745 \\
2023-Q3 & +0.0167 & 15\%/71\%/14\% & 48\% & 0.576 & 0.750 \\
2023-Q4 & +0.0328 & 26\%/0\%/74\% & 46\% & 0.613 & 0.737 \\
2024-Q1 & +0.0219 & 56\%/39\%/5\% & 50\% & 0.564 & 0.766 \\
2024-Q2 & +0.0209 & 81\%/15\%/4\% & 58\% & 0.477 & 0.803 \\
2024-Q3 & +0.0107 & 86\%/0\%/14\% & 54\% & 0.517 & 0.782 \\
2024-Q4 & +0.0663 & 13\%/27\%/61\% & 45\% & 0.618 & 0.730 \\
\midrule
Mean & $\mathbf{+0.0252}$ & --- & --- & --- & --- \\
\bottomrule
\end{tabular}
\end{table}

\end{document}